\documentstyle[prl,floats,epsbox,twocolumn,aps]{revtex}

\begin{document}


\twocolumn[\hsize\textwidth\columnwidth\hsize\csname
@twocolumnfalse\endcsname
\title{\bf The Ground State of the Two-Dimensional Hubbard Model}
\author{Yoshihiro Asai}
\address{Physical Science Division, Electrotechnical Laboratory (ETL), \\
Agency of Industrial Science and Technology (AIST), \\
Umezono 1-1-4, Tsukuba, Ibaraki 305-8568, Japan}
\date{Received on \today}
\maketitle

\widetext
\begin{abstract}
We have studied the ground state of the two-dimensional Hubbard
model by using the recently proposed adaptive sampling quantum monte
carlo (ASQMC) method. We have paid attention to the model whose
non-interacting band dispersion is almost flat near $(\pi,0)$.
To minimize the effect of the finite size gap overlying on the Fermi
level, we have tuned both filling and band structure.
We found enhancement of the d-wave correlation function at large
distance, the spin gap and the momentum distribution function 
consistent with the d-wave gap. We also found the coexistence of both
the commensurate and incommensurate peaks in $S(\vec{q})$,
which does not contradict a recent experimental finding that
both the resonance peak and the incommensurate peaks reside
in the same doping level of YBCO and BSCCO.
\end{abstract}


\pacs{}

]

\narrowtext

It is very difficult to calculate ground state properties of large 
two-dimensional (2D) Hubbard clusters away from half filling when $U/t$
is not small because of the notorious negative sign problem for quantum
monte carlo (QMC) methods. We have recently proposed the adaptive sampling
quantum monte carlo (ASQMC) method to cope with the problem.~\cite{Asai1}
The start of the project to develop the ASQMC method was influenced by
the publication of the constrained path (CP) monte carlo method
by Zhang, Carlson and Gubernatis (ZCG).~\cite{ZCG}
However we found it is possible to reduce the negative sign ratio without
any constraint if we adopt a new update scheme.
This may be possible because the update scheme adopted by ZCG itself seems
to suppress the negative sign ratio, which should be the reason why their
method is superior in accuracy than the more elaborate positive projection
method proposed by Fahy and Hammann 8 years ago.~\cite{FH}
We developed our update scheme within the frame work of the "exact
update method" of the auxiliary field quantum monte carlo method.
Our scheme is not only easier to understand for those who have been
working on Hubbard and related models but also it is superior in accuracy
than the ZCG's scheme because it is free from the inaccurate mixed
estimator method for measurements and its variants which are indispensable
to the CP method.
This should be one of the main reasons why our ASQMC method gives
more accurate results than the CP method. 
Actually, we observed the ASQMC gives better value for the ground state
energy of 14 electron $4 \times 4$ cluster when $U/t=12$.~\cite{Asai1}
Even without constraint, the negative sign ratio is reduced largely in the
all cases studied, which makes measurements stable. $<Sign>$ decrease to 
$0$ very slowly as we increase the projecting imaginary time $\tau$.
(Fixing $\Delta \tau$ and $\tau_c$ constant.) 
Typically, $\tau=20$ to $30$ is feasible. We have made the serial
correlation test in many cases and found that samples obtained by the
ASQMC are statistically independent and the variance calculated by the
standard formula gives a good estimate for the true variance. As an
example we show the autocorrelation function for the d-wave
superconducting correlation function at $\vec{q}=0$ in the case of
parameter set (i) described below in Fig. \ref{fig:autcorr}.

\begin{figure}[htbp]
\epsfxsize=8.5cm
\epsffile{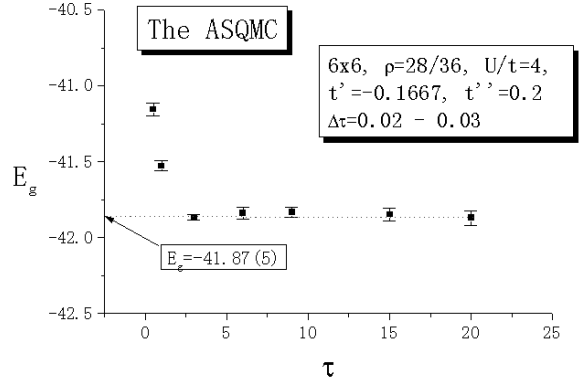}
\vspace*{0.5cm}
\caption{The projecting imaginary time $\tau$ dependence of the ground
state energy of the $6 \times 6$ cluster.}
\label{fig:convg}
\end{figure}

\begin{figure}[htbp]
\epsfxsize=8.5cm
\epsffile{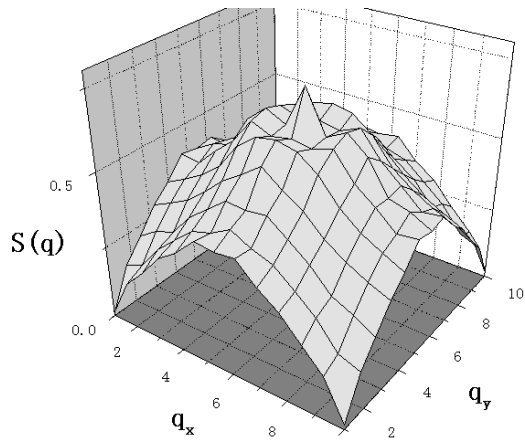}
\vspace*{0.5cm}
\caption{The spin correlation function $S(\vec{q})$ of the $12 \times
12$ cluster.}                         
\label{fig:spincorr} 
\end{figure}

We used the ASQMC method to study the ground state properties
of the 2D Hubbard model.
The parameter region we paid our attention was such that
the non-interacting band dispersion was almost flat near the
$(\pi,0)$ point. We tuned electron filling and band dispersion
such that the energy difference between the highest occupied level (HOL)
and the lowest unoccupied level (LUL) was close to zero.
The Hubbard model we studied may belong to different
class than that frequently studied on the bipartite lattice. 
We have studied various parameter regions of $6 \times 6$,
$8 \times 8$, $10 \times 10$ and $12 \times 12$ clusters.~\cite{Asai2}
Here we discuss following three cases;
(i) $6 \times 6$ lattice with $t' = -0.1667$, and $t''= 0.2$. $U=4$.
The number of electron is $28$. 
(ii) $10 \times 10$ lattice with $t' = -0.2$, and $t'' = 0.05$. $U=4$. 
The number of electron is $84$.
(iii) $12 \times 12$ lattice with $t' = -0.1676$, and $t''= 0.2$. $U=4$. 
The number of electron is $92$.
To make sure that our result is convergent to the $\tau \to \infty$ limit,
we show the $\tau$ dependence of the ground state energy of (i) in 
Fig. \ref{fig:convg}.
The calculation was stable at least up to $\tau=20$, but the energy
converged at $\tau=3$. We observed similar fast 
convergence in the case of (ii). Hence, we proceeded calculations
with $\tau=3$ to save CPU time.
The fast convergence suggests that energy gaps among the states with the
same symmetry are relatively large, while the gap between the ground and
low lying excited states in the finite size system can be small, if these
state have different symmetries. This is because we are using the
projector method.
We found the enhancement of the d-wave superconducting correlation
at large distance in the ground state when we used the parameter sets (i)
and (ii). We also found the momentum distribution function can be well
described by using the d-wave BCS mean field theory.
We found exponential decay of the distance dependence of the spin
correlation function, which is suggestive of opening the 
spin gap.~\cite{Asai2}
All these are consistent with the d-wave superconducting ground
state. Even more surprisingly, the commensurate peak coexists with the
incommensurate peaks in $S(\vec{q})=\int d \omega S(\vec{q}, \omega)$ of
the superconducting state. 
This is characteristic to the parameter region involving the three
parameter sets. The coexistence was most clearly seen in a over doped
region (iii) which was shown in Fig. \ref{fig:spincorr}.
(The peak interval between the incommensurate peaks is largest in (iii).)
Experiments on YBCO and BSCCO seem to suggest that both the resonance peak
and the incommensurate peaks exist at different $\omega$ but in the same 
filling.~\cite{Mook}
{\it The suggestion from experiments does not contradict our result
on $S(\vec{q})$.}
In the other parameter regions, we did not so far observe similar
$S(\vec{q})$. 
The numerical results so far obtained by using the ASQMC method suggest
that the band dispersion close to $(\pi,0)$ seems to be very crucial to
the ground state of the 2D Hubbard model.
%


%
%

%
%
%

%
%
%

\begin{figure}[htbp]
\epsfxsize=8.5cm
\epsffile{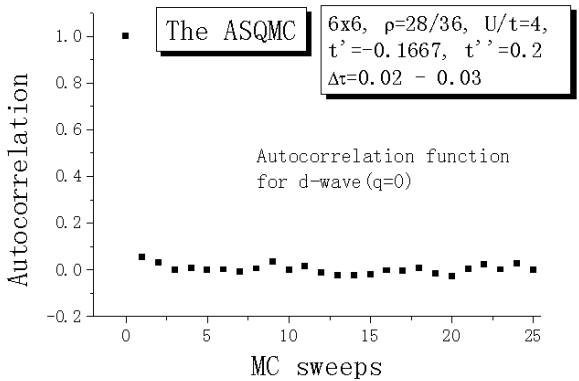}
\vspace*{0.5cm}
\caption{Autocorrelation function for the d-wave superconducting
correlation at $\vec{q}=0$ in the case of the parameter set (i).}
\label{fig:autcorr}
\end{figure}

\end{document}